# Sensitivity Analysis of a Graphene Field-Effect Transistors by means of Design of Experiments

Giovanni Spinelli[1*], Patrizia Lamberti[1], Vincenzo Tucci[1],
Francisco Pasadas[2], David Jiménez [2]

[1] *Department of Information and Electrical Engineering and Applied Mathematics*
*University of Salerno, Via Giovanni Paolo II, Fisciano (SA) ITALY*
[2] *Departament d'Enginyeria Electrònica, Escola d'Enginyeria,*
*Campus de la UAB, 08193 Bellaterra (Cerdanyola del Vallès) Barcelona, Spain*

* Corresponding authors. Fax:+39 089 964218 E-mail address: gspinelli@unisa.it;

**Abstract.** Graphene, due to its unique electronic structure favoring high carrier mobility, is considered a promising material for use in high-speed electronic devices in the post-silicon electronic era. For this reason, experimental research on graphene-based field-effect transistors (GFETs) has rapidly increased in the last years. However, despite the continuous progress in the optimization of such devices many critical issues remain to be solved such as their reproducibility and performance uniformity against possible variations originated by the manufacturing processes or the operating conditions. In the present work, changes of the $I_D$-$V_{DS}$ characteristics of a Graphene Field-Effect Transistors, caused by a tolerance of 10% in the active channel (i.e. its length and width) and in the top oxide thickness are numerically investigated in order to assess the reliability of such devices. Design of Experiments (DoE) is adopted with the aim to identify the most influential factors on the electrical performance of the device, so that the fabrication process may be suitably optimized.

## 1. Introduction

Graphene, a material obtained by an exfoliation process from graphite  is a two-dimensional (2D) thin crystalline film formed by sp2-hybridized carbon atoms arranged in a honeycomb structure [1]. Due to this particular molecular structure, graphene shows remarkable mechanical properties, such as aYoung's modulus of 1 TPa and a tensile strength of 130 GPa for single layer [2]. Graphene has been adopted in different applications ranging from sensors, actuators, energy storage, biomedical aids to name a few. As an example, flexible biosensors have been proposed because of the atom-thick two-dimensional conjugated structures and large specific surface areas, which are favorable elements, especially if properly functionalized, for capturing small molecules [3]. More recently, due to its extremely high carrier mobilities for both electrons and holes (as high as $10^4$ $cm^2/V{\cdot}s$ at room temperature), tunable band gap and unique electronic structure considerably different from that of materials conventionally employed in solid-state electronics, graphene-derived nanomaterials are emerging as promising candidates for post-silicon electronics devices with high performances [4, 5, 6]. Although graphene-based field effect transistors (GFETs) require planar manufacturing processes similar to those already present in the existing technology, they could be more convenient compared to conventional semiconductors, like germanium and silicon, due to the mass-scalability, improved performances and price effectiveness of the resulting devices. A gate, a channel-region connecting source and drain electrodes, and a dielectric material separating the gate from the channel are the main electrical parts of a FET. The basic functioning of such device requires the modulation of the channel conductivity by means of a voltage applied between the gate and source (i.e. VGS), and thus on the control of the drain current collected to the source terminal due to a voltage (i.e. VDS) applied between these two last electrodes. In a GFET, a suitable graphene sheet acts as active channel. Advantages beyond conventional metal–oxide–semiconductor but also performance limits of these emerging electronic devices when exploited for both digital and analog applications are discussed in a review focused on electronics based on two-dimensional materials [7]. A new generation of vertical field-effect transistors based on heterostructures of graphene and tungsten disulphide has been proposed for flexible and transparent electronics [8]. However, despite the achieved results, many critical issues remain to be solved. In particular, especially for electronic devices, it is required to ensure stable performances and proper functioning within certain limits even in the presence of unavoidably variations of either the constituent physical parameters or the operating conditions.

In order to analyze such aspects, a simulation study of graphene nanoribbon based field-effect transistors (GNR-FETs), by solving 3D Poisson and Schrödinger equations, has been performed. The electrical behaviour of such devices and their sensitivity against variability of channel chirality and leakage problems due to band-to-band tunneling has been investigated [9]. In the present work, Design of Experiments (DoE) approach is adopted for the optimization of a large-signal model of GFETs. In particular, an assessment of its reliability is carried out due to a tolerance of 10% in the fabrication process of the active channel (i.e. its length and width) and of the top oxide thickness. Moreover, the most influential factor among them is identified.

## 2. Models and methods

### *2.1 Graphene field-effect transistor (G-FET)*

Among the different typical configurations used in GFETs, we consider the so-called dual-gate solution in which the control of free carrier concentration in the channel can be obtained by both gate biases. It is worth noting (Fig. 1) that, in addition to the main electrodes configuration, some important manufacturing process parameters strongly influence the overall performance of the device, such as width (*W*) and length (*L*) of the graphene-based channel and the top oxide thickness ($L_t$).

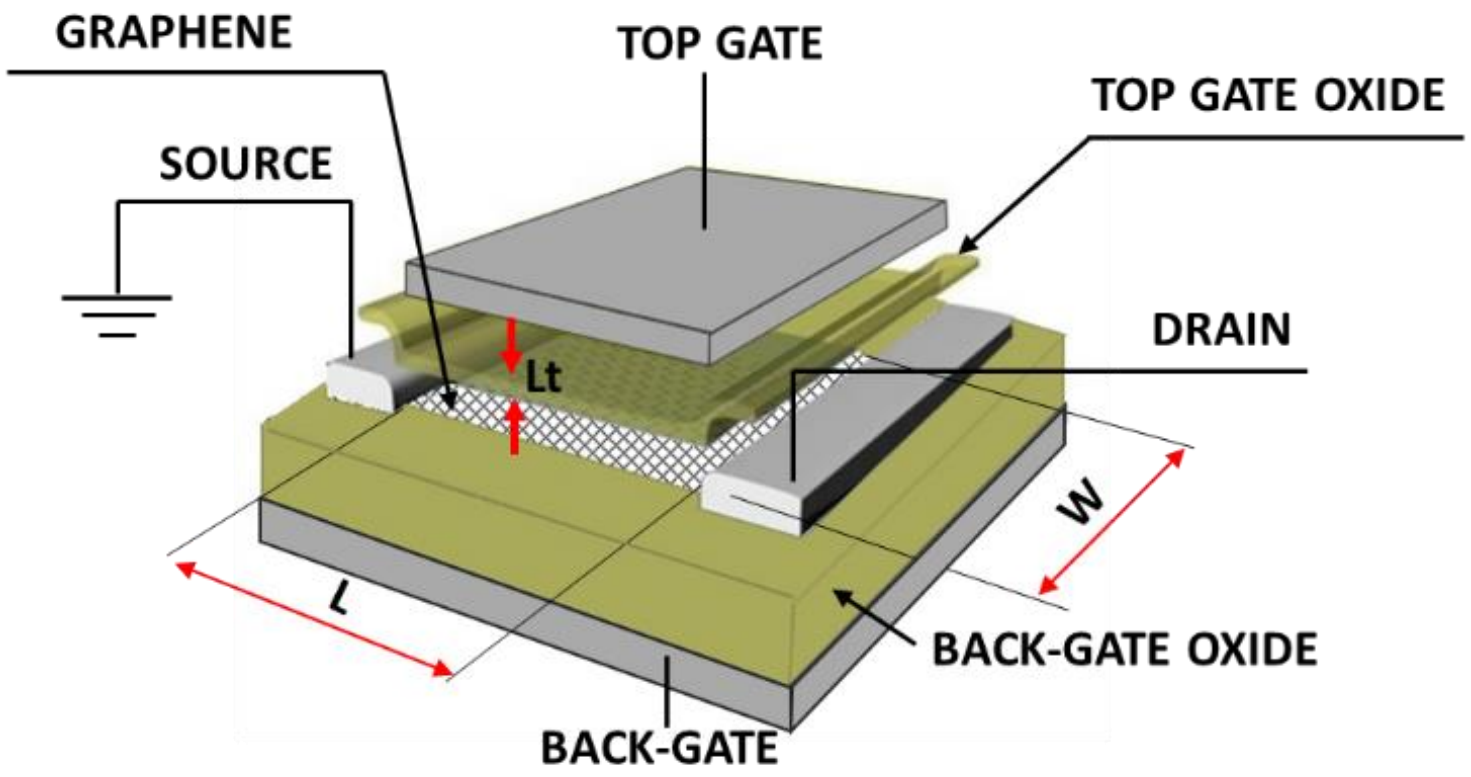

**Fig. 1:** 3D schematic of the GFET of the component implemented in Cadence Virtuoso environment. A graphene sheet plays the role of the active channel. The source is grounded and considered as the reference potential in the device.

A design leading to performances insensitive to the tolerances affecting these parameters is desired. Different simulations have been performed with Cadence Virtuoso Spectre Circuit Simulator by using an equivalent electric model of the component (Fig.2), The model has been developed and validated with experimental data by some of the authors in previous studies [10, 11]. For completeness and clarity reasons, the main features of this component are illustrated in the sequel. For more details the reader may refer to the mentioned papers [10, 11].

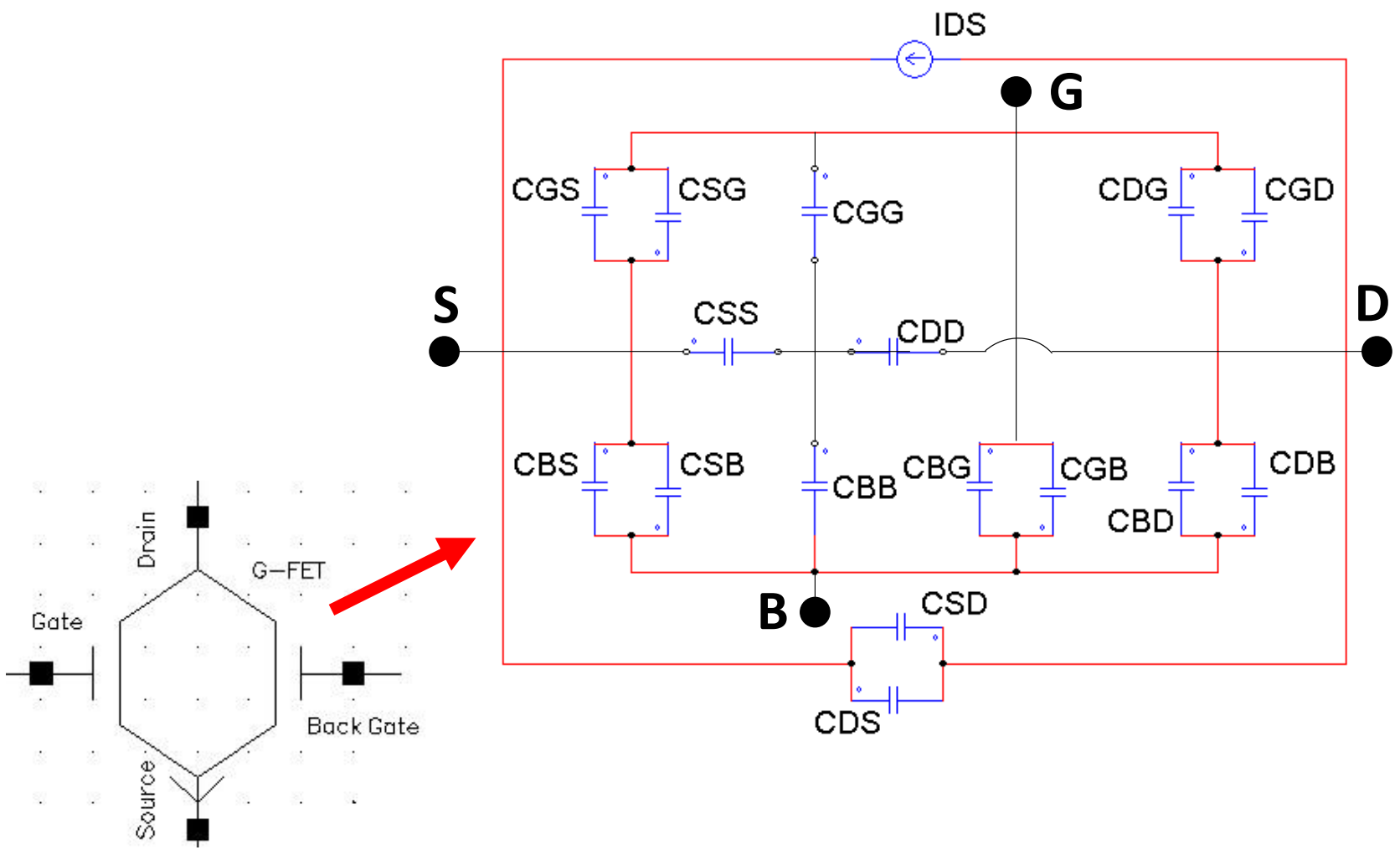

**Fig. 2:** Electrical symbol of a G-FET implemented in Cadence Virtuoso environment and its equivalent circuit formed by the drain current model and the intrinsic capacitance model.

A four-terminal FET can be modeled with a charge-based capacitance model including 16 capacitances in total, 4 of which are self-capacitances and 12 intrinsic trans-capacitances. These capacitances can be collected in a matrix *M* as follows:

$$M = \begin{bmatrix} Cgg & -Cgd & -Cgs & -Cgb \\ -Cdg & Cdd & -Cds & -Cdb \\ -Csg & -Csd & Css & -Csb \\ -Cbg & -Cbd & -Cbs & Cbb \end{bmatrix} \quad (1)$$

where each element *Cij* accounts for the dependence of the charge at terminal *i* with respect to a varying voltage applied to terminal *j* when the voltage at the remaining terminals is maintained constant:

$$C_{ij} = -\left.\frac{\partial Q_i}{\partial Q_j}\right|_{i\neq j} \;;\quad C_{ij} = \left.\frac{\partial Q_i}{\partial Q_j}\right|_{i=j} \quad (2)$$

where *i* and *j* stand for *g*, *b*, *d* and *s* which are the top gate, bottom gate, drain and source electrodes.

It is worth pointing out that only 9 of the 16 intrinsic capacitances are independent since each row, as well as each column of M must sum to zero so that the device is charge-conservative, thus fulfilling Kirchhoff's current law at its electrodes. Therefore:

$$\sum_{i=1}^{4}\sum_{j=1}^{4} C_{ij} = 0\,;\quad \sum_{j=1}^{4}\sum_{i=1}^{4} C_{ij} = 0 \quad (3)$$

The drain current can be calculated, based on drift-diffusion (DD) theory, by the following expression:

$$I_{ds} = \mu \frac{W}{L} \int_{V_{CS}}^{V_{CD}} Q_{tot}(V_c) \frac{dV}{dV_c} dV_c \quad (4)$$

where $\mu$ is the effective carrier mobility for both electrons and holes, $Q_{tot}$ is the transport sheet charge density depending on the $V_c$ that is the voltage drop across the graphene layer. The integration extremes are $V_{cs} = Vc|_{V=0}$ and $V_{CD} = Vc|_{V=V_{ds}}$ whereas the term $dV/dV_c$ is calculated as:

$$\frac{dV}{dV_c} = 1 + \frac{C_q(V_C)}{C_t + C_b} \quad (5)$$

with $C_t$, $C_b$ and $C_q$, top and bottom oxide capacitance and quantum capacitance, respectively [10].

The most relevant parameters adopted for the GFET during the numerical simulation performed in this study are reported in Table I.

**Table I** Input parameters of the GFET used to simulate the output characteristics.

| Parameter | Value | Parameter | Value |
|---|---|---|---|
| $T$ | 300 K | $\varepsilon_{bottom}$ | 3.9 |
| $\mu$ | 4500 cm$^2$/Vs | $L$ | 500 nm |
| $V_{gs0}$ | 0.613 V | $W$ | 30 µm |
| $\Delta$ | 0.095 eV | $L_t$ | 4 nm |
| $\varepsilon_{top}$ | 12 | | |

More in details, $T$ is the temperature, $\mu$ represents the effective carrier mobility; $V_{gs0}$ is the top gate voltage offset; $\Delta$ is the inhomogeneity of the electrostatic potential due to electron-hole puddles; $\varepsilon_{top}$ and $\varepsilon_{bottom}$ refer the top and bottom oxide relative permittivity, respectively.

### *2.2 Design of Experiments (DoE): procedure for sensitivity analysis of the device*

A system, whether it is a product or a process, is robust if it performs properly in a wide range of conditions. Design of Experiments (DoE) is a useful method for evaluating the relationship between factors conditioning a process and the output of that process. As schematized in Fig. 3, at design stage, there are some controllable factors (X variables) and others not controllable (N, noise variables) including environmental variables, tolerances on X variables as well as unknown noise sources.

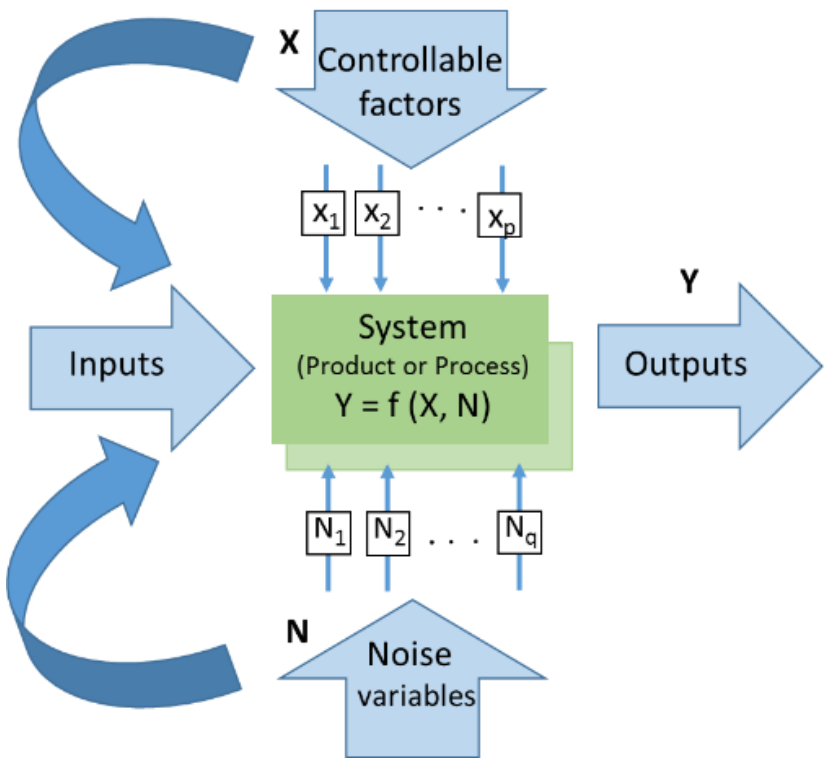

**Fig. 3:** Schematic representation of a system at the design stage.

Variation of controllable factors (**X=(**$x_1$, $x_2$,..,$x_p$**)**) combined with variation of noise variables (**N=(**$N_1$, $N_2$,..$N_q$**)**) impact on the transfer function f(**X**, **N**) thus affecting the output variables (**Y=(**$y_1$,$y_2$,…,$y_m$**)**). The factor, which is the most influential on an output variable $y_k$, with $k = 1, \ldots m$, can be identified by means of DoE. Moreover, the controllable input factors can be adequately selected to optimize the output or to limit the impact on the results by the external factors such as their variation (robust design, RD) [12]. In the present work, the variations of the output characteristics of the device, caused by tolerances in the fabrication process of the active channel (i.e. its length L and width W) and of the top oxide thickness (i.e. $L_t$) are investigated as input controllable factors. A tolerance of 10% is considered, thus obtaining the maximum and minimum values with respect to the nominal ones summarized in Table II. Matlab® routine are specifically developed in order to carry out DoE.

**Table II** Nominal, minimum and maximum values for the input variables due to a 10% of tolerance.

| Input Variable | Nominal Value | Minimum Value | Maximum Value |
|---|---|---|---|
| L | 500 nm | 450 nm | 550 nm |
| W | 30 μm | 27 μm | 33 μm |
| $L_t$ | 4 nm | 3.60 nm | 4.40 nm |

It is worth evidencing that the possible values of the variable parameters are uniformly distributed in the considered ranges. We have to consider a set of 3 parameters $x_i$, with i=1,…,3, describing the variable inputs of the device, each of them varying in a well-defined range, i.e. $x_i = [x_{i_min}, x_{i_max}] \epsilon \mathcal{R}$. The corresponding parameters space is the compact $D$ defined as: $\mathcal{D} = x_1 \times x_2 \times x_3 \subset \Re^3$. The outputs of the system can be computed for each set of input parameters $\underline{x} = (x_1, x_2, x_3) \epsilon D$. If the three-level selection is adopted for each factor $(x_{i_min}, x_{i_nom}, x_{i_max})$, a full factorial array approach leads to consider $3^3$=27 points $\epsilon$ $D^* \subset D$ that generate the scattered data of the responses required for analyzing the sensitivity to each factor [12].

The nominal solution $y_{k_nom} = (x_{1_nom}, x_{2_nom}, x_{3_nom})$falls in one of these computed data used to evaluate the main effect of each factor and to select the most influencing one.

Moreover, inside these ranges DoE is used to derive the estimation of the bounding $Y_k$ (i.e. $Y^*_k$) of the particular system performance $y_k$, where:

$$Y_k^* = \left[\min_{\underline{x}\in D^*} y_k\, , \max_{\underline{x}\in D^*} y_k\right] \subseteq\ Y_k = \left\{y_k \middle| \underline{x} \in D\right\} \tag{6}$$

In particular, it is possible to obtain an overestimation of the minimum (lower bound) and an underestimation of the maximum (upper bound):

$$\min_{\underline{x}\in D^*} y_k \geq \min_{\underline{x}\in D} y_k \quad ; \quad \max_{\underline{x}\in D^*} y_k \leq \max_{\underline{x}\in D} y_k. \tag{7}$$

Furthermore, their difference can be used as a measure of the sensitivity of the particular system performance ($S_{Yk}$) with respect to the parameter variations:

$$S_{Yk} = \frac{\max_{\underline{x}\in D^*} y_k - \min_{\underline{x}\in D^*} y_k}{y_{k_nom}} \times 100 \tag{8}$$

Lower the value of eq. (8), lower is the sensitivity [13, 14].

## 3. Results and discussion

### *3.1 Variability analysis of the output characteristics*

Fig. 4 shows the voltage sources VGS and VDS applied to polarize a GFET and the direction assumed for the terminal currents (namely $I_D$, $I_G$, and $I_S$). The conceptual circuit, schematized in Cadence environment, is classically adopted to measure the output characteristics, which are a family of curves, each one obtained for a fixed value of VGS.

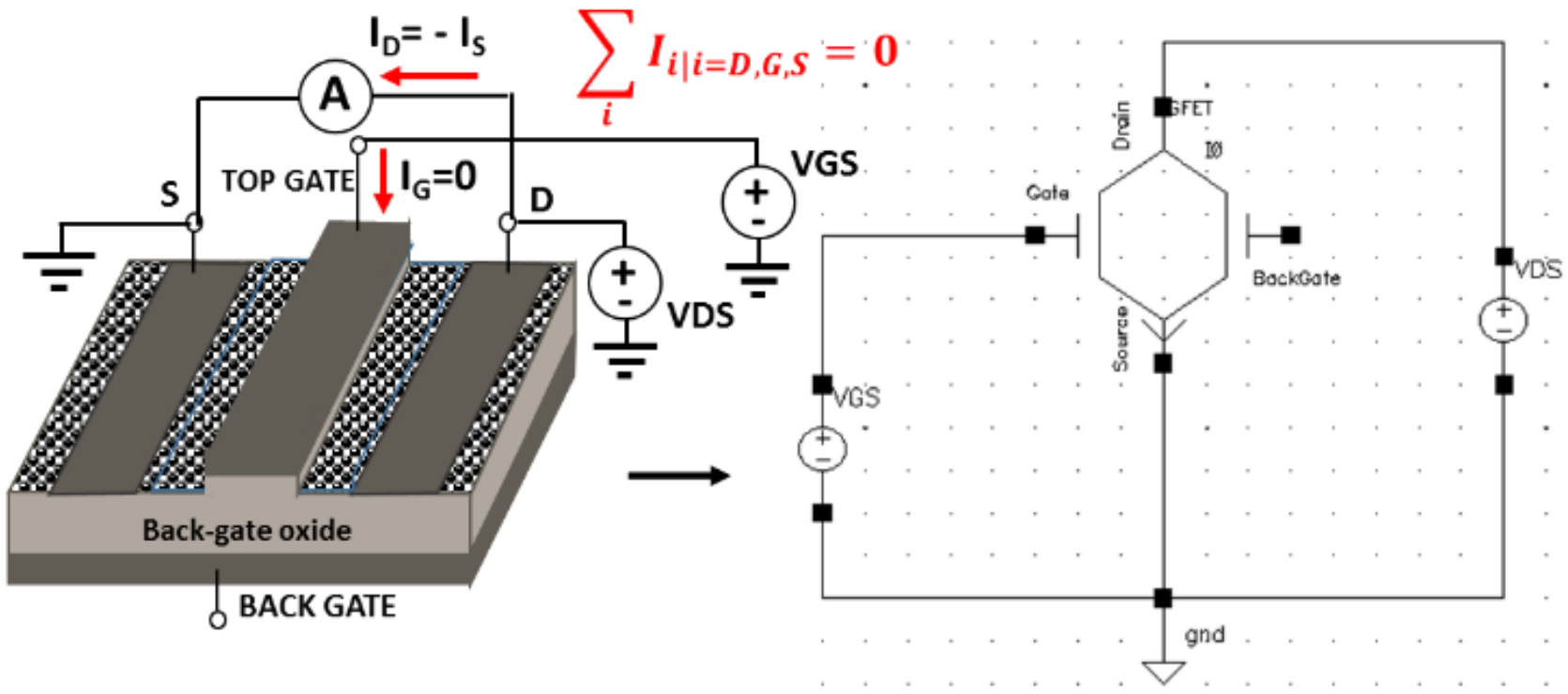

***Fig.4*** *Polarization of a GFET with current directions and relative electrical schematic circuit implemented in Cadence.*

Fig. 5 shows the influence of each variable parameter on the simulated output characteristics performed with the "one factor at a time" approach instead of varying multiple factors simultaneously. In particular, a single parameter ranges between the minimum and the maximum value whereas all other are fixed to their nominal ones. Tolerances in the length $L$ and width $W$ of the channel, as well as in the top-gate thickness $L_t$ are investigated in Fig. 5a), 5b) and 5c), respectively, when a $V_{GS}$ of 1V is applied. From Fig. 5a), it is interesting to note that a tolerance of 10% in the channel length ($L$) does not affect much the output characteristics or more specifically the drain current that flows towards the source electrode of the device. This is probably due to the fact that the channel resistance is comparable to the contact resistances, which are taken into account by the developed equivalent electric model for the GFET [15]. An increase in the current $I_D$ with the increasing of the channel width ($W$) is observed due to an increment of the charge carriers (Fig. 5b)). As expected, an influence of the top-gate thickness is found, given the validity of the following relationships:

$$I_D \propto \mu \cdot C_{ox} \text{ and } C_{ox} \propto \frac{\varepsilon_{oxide}}{Thickness\ of\ oxide} \quad (9)$$

Therefore, coherently to the eq. (6) and as shown in Fig. 5c), an increase in the oxide layer causes a decrease in the drain current.

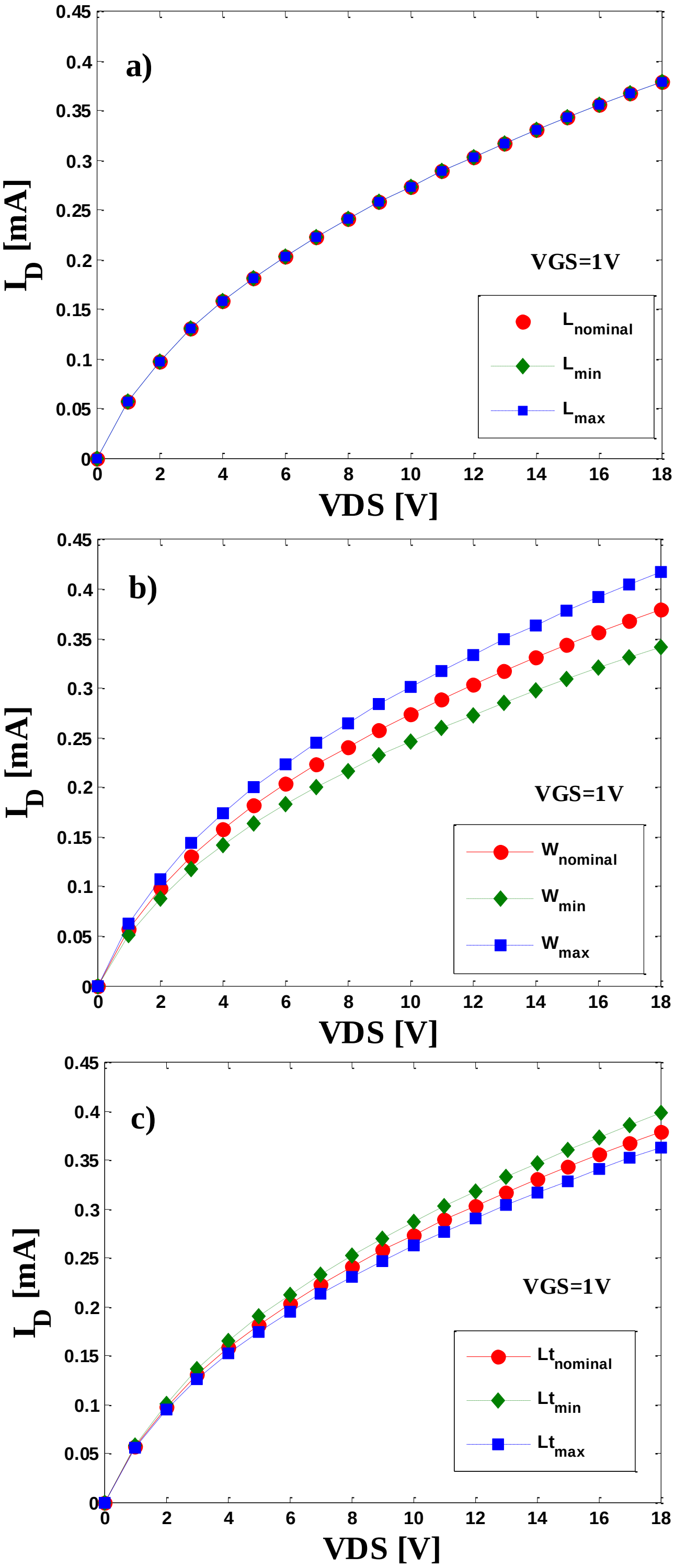

**Fig. 5:** Influence of the tolerance of the parameters L, W and Lt (in a), b) and c) respectively) on the output characteristics evaluated for VGS fixed at 1V.

Fig.6 reports the nominal solution and the bounding of the drain current IDS defined according to the eq. (6) and obtained by means of three level full factory discretization of the data summarized in Table II. For low polarization value (VDS=1V) the drain current varies up to 25% with respect to its nominal value caused by a combination of 10% of tolerance on each factor whereas at highest voltage polarization (VDS=18V) it achieves about 30% of variation. The latter is approximately equal to the sum of a negligible, 20% and 9% of variation provided by $L$, $W$ and $Lt$ conditioning factor, respectively, investigated in Fig.5.

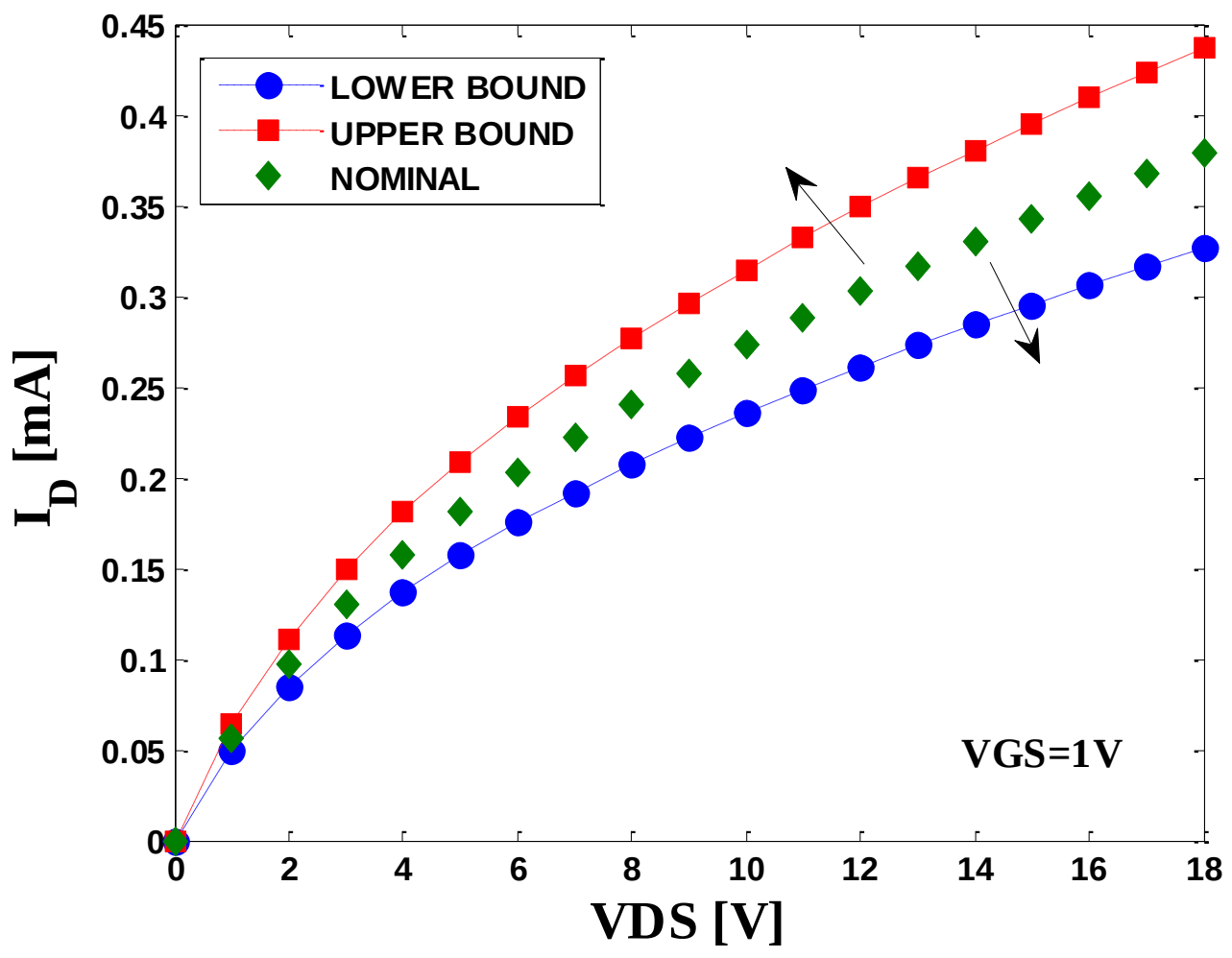

**Fig.6:** Upper and lower bound for the IDS vs VDS obtained taking into account variation of the factors L, W and Lt of Tab. II.

### *3.2 DoE Results: Dex Scatter Plot and Main factor Plot*

Dex Scatter Plot (DSP) and Main factor Plot (MfP) are typically graphics used for the representation and subsequent interpretation of the collected data. The output of the system, i.e. the $I_D$ current as function of the VDS, is computed for 27 points $x = (L, W, L_t)\epsilon D$ generating the scattered data of the response on which the analysis of the impact of each factor is performed. In Fig. 6 the DSP for $I_D$ current computed for VGS=1V and with VDS=10V is reported. These data are represented in column by fixing one factor at time and varying the remaining two in the adopted variability range given in Tab II. This means that the first, second and third red-point columns reported in Fig. 7 are given by fixing $L$ parameter to its minimum, nominal and maximum value respectively, whereas $W$ and $L_t$ vary in their ranges. If the same 27 points are organized by fixing $W$ to its minimum, nominal and maximum value respectively and by considering $L$ and $L_t$ factors variable in their ranges, the blue-squared columns are obtained. Finally, the green-diamond columns represent the DSP od the output $I_D$ for the $L_t$ factor.

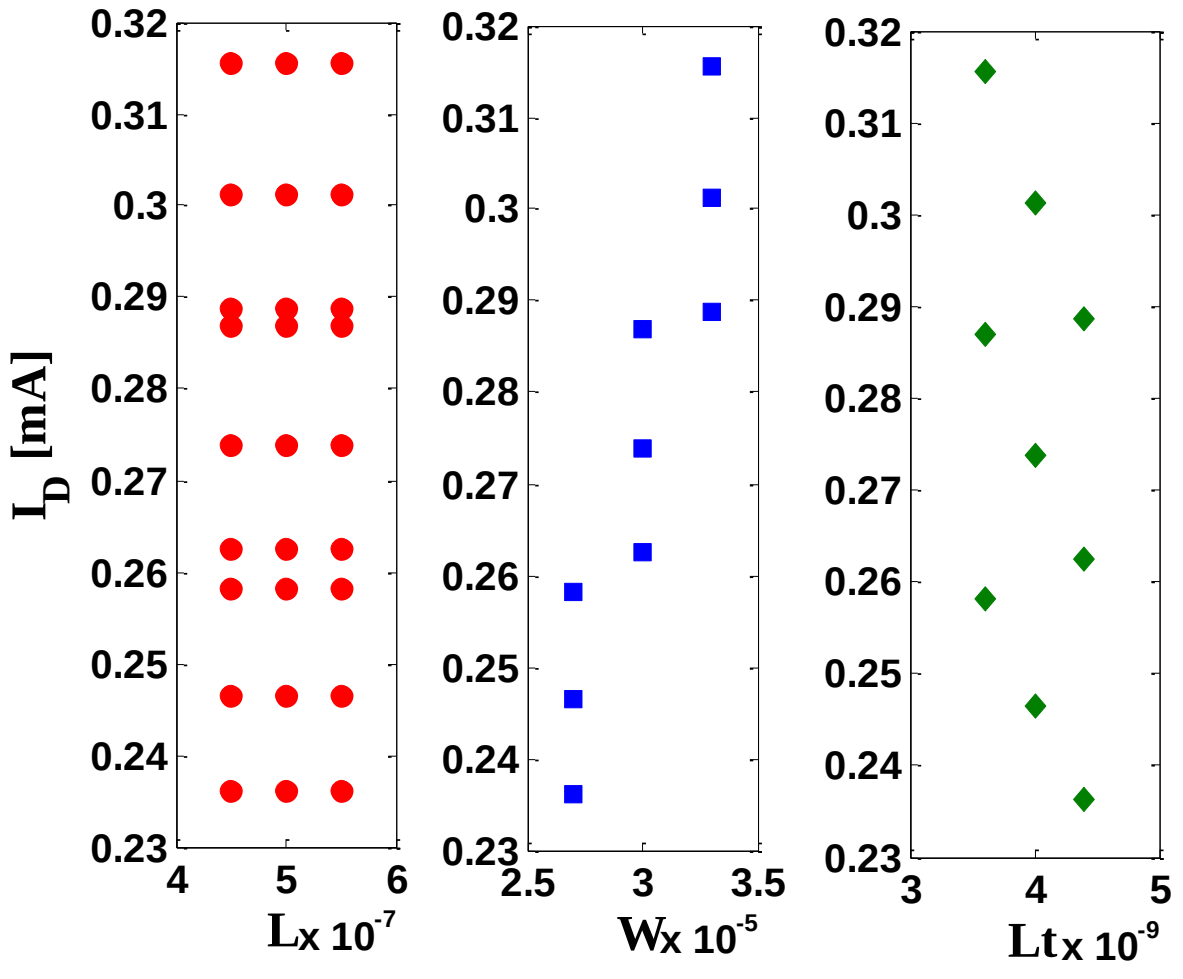

**Fig. 7:** DSP of the drain current computed for VGS=1V and VDS=10V with respect factors L, W and Lt of Tab. II

As it possible to observe from Fig. 7, the higher columns are in the DSP for *L* factor. These columns are aligned and of the same width, indicating that the bounding of the drain current is not influenced by the variation of the length of the active channel L. Conversely, the shortest and less aligned columns are in DSP for *W* factor, indicating that the variability of the width of the active channel is responsible of the drain current variation. These columns are also with increasing value for higher considered W. Finally, DSP for $L_t$ factor represented with diamond-green markers show that also $L_t$ introduce effect on $I_D$ but lower than that of *W* and, moreover, it lows the drain current values. All these informations in terms of signum and intensity of dependence can be better collected by using the MfP of Fig. 8.

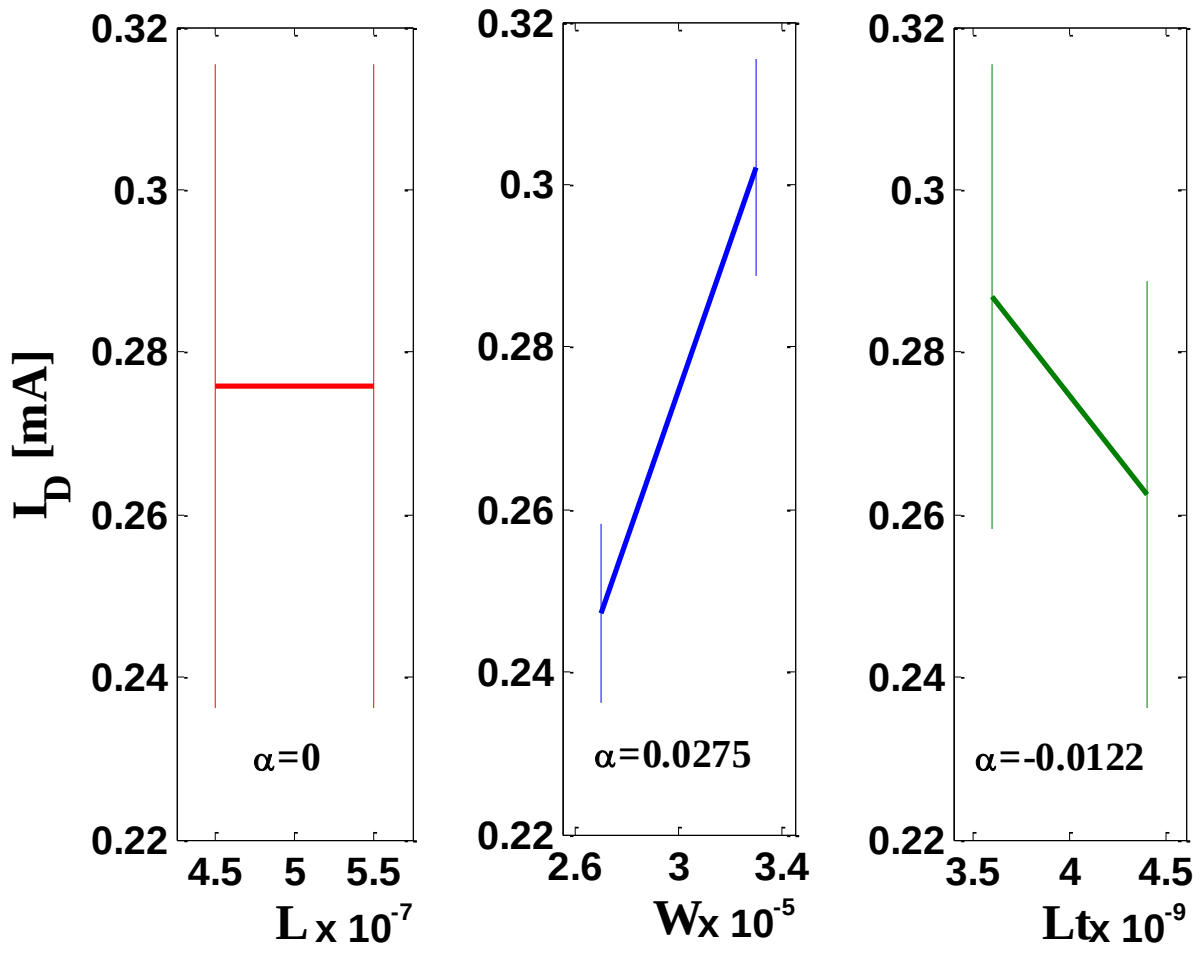

**Fig.8:** MfP of the drain current computed for VGS=1V and VDS=10V with respect factors L, W and Lt of Tab. II

The MfP is obtained from the DSP by considering the excursion at lower and higher level for each parameter and by graphing the segment between the corresponding mean values. The slope of this straight-line quantifies the influence of the particular design parameter on the considered output. A null value is indicative of an independence; a positive/negative coefficient provides information about the dependence direction (increasing/decreasing, respectively). The higher the value, the greater the dependency. From the analysis of such graphical representations in Fig. 8, it is possible to extract a coefficient α=0, 0,0275 and -0.0122 that is indicative of an independence from *L*, a positive dependence from *W* and a negative dependence from $L_t$ respectively, with *W* the most influencing one exhibiting the maximum α value.

## 4. Conclusions

Variations of the output characteristics of a GFET, caused by 10% of tolerance in the fabrication process of the active channel and of the top oxide thickness are investigated by adopting DoE approach. In particular, the width W of the channel shows the greater impact on the response and therefore, has to be suitably optimized in order to achieve the desired performances of the G-FET devices. More in details, an overall variation of 30% for the drain current (when a VDS=18V is applied) is observed with the remarkable contribution of 20% due to the uncertainty of width channel. Future works will attempt to analyze other performance functions, (i.e. cut-off frequency, delay time) and to explore other influential factors.

## Acknowledgments

This work has been supported from H2020-SGA-FET- Graphene Flagship- Graphene Core 2, G.A.: 785219 and 696656, and by the Ministerio de Economía y Competitividad under Grant TEC2015-67462-C2-1-R.